\documentclass[aps,prb,twocolumn,superscriptaddress,showpacs]{revtex4}
\usepackage{bm}
\usepackage{amsmath}
\usepackage{amssymb}
\usepackage{textcomp}
\usepackage{graphicx}
\usepackage{color}

\usepackage{soul,xcolor}
\setstcolor{red}

\usepackage{hyperref}
\newcommand*{\doi}[1]{\href{https://doi.org/\detokenize{#1}}{doi: \detokenize{#1}}}

\input{epsf}

\begin{document}

\title{Mechanically Induced Thermal Breakdown in Magnetic Shuttle Structures}

\author{O. A. Ilinskaya}
\email{ilinskaya@ilt.kharkov.ua} \affiliation{B. Verkin Institute
for Low Temperature Physics and Engineering of the National
Academy of Sciences of Ukraine, 47 Nauki Ave., Kharkiv 61103,
Ukraine}
\author{S. I. Kulinich}
\affiliation{B. Verkin Institute for Low Temperature Physics and
Engineering of the National Academy of Sciences of Ukraine, 47
Nauki Ave., Kharkiv 61103, Ukraine}
\author{I. V. Krive}
\affiliation{B. Verkin Institute for Low Temperature Physics and
Engineering of the National Academy of Sciences of Ukraine, 47
Nauki Ave., Kharkiv 61103, Ukraine}  \affiliation{Physical
Department, V.N. Karazin National University, Kharkiv 61077,
Ukraine}
\author{R. I. Shekhter}
\affiliation{Department of Physics, University of Gothenburg,
SE-412 96 G{\" o}teborg, Sweden}
\author{H. C. Park}
\email{hcpark@ibs.re.kr}
\affiliation{Center for Theoretical Physics of Complex Systems,
Institute for Basic Science (IBS), Daejeon 34051, Republic of
Korea}
\author{M. Jonson}
\affiliation{Department of Physics, University of Gothenburg,
SE-412 96 G{\" o}teborg, Sweden}

\date{\today}

\begin{abstract}
A theory of a thermally induced single-electron ``shuttling"
instability in a magnetic nano-mechanical device subject to an external magnetic field is presented in the Coulomb blockade regime of electron transport.
The model magnetic shuttle device considered comprises a movable metallic grain suspended between two magnetic leads, which are kept at different temperatures
and assumed to be fully spin polarized with antiparallel magnetizations. For a given temperature difference shuttling is found to occur for a region of external magnetic
fields between a lower and an upper critical field strength, which separate the shuttling regime from normal small-amplitude ``vibronic" regimes. We find that
(i) the upper critical magnetic field saturates to a constant value in the high temperature limit and that the shuttle instability domain expands with a decrease of the temperature;
(ii) the lower critical magnetic field depends not only on the temperature independent phenomenological friction coefficient used in the model but also on intrinsic friction
(which vanishes in the high temperature limit) caused by magnetic exchange forces and electron tunneling between the quantum dot and the leads.
 The feasibility of using thermally driven magnetic shuttle systems to harvest thermal breakdown phenomena is discussed.
 \end{abstract}

\maketitle

\section{Introduction}

Mechanically promoted electric transport, being one of the most interesting
features of nanoelectromechanics (NEM), offers a new functionality to devices
on the nanometer length scale. Shuttling of electrons, as predicted in Ref.~\onlinecite{shekhter}
and actively studied both theoretically and experimentally, is a prominent example
of this statement (see, e.g., the review [\onlinecite{review}]).

Heat transport in nanostructures is a subject of enhanced
interest \cite{benenti} especially due to the importance of heat
removal on a nanometer length scale. Electrically induced
mechanical shuttling of electrons results in an exponential
decrease of electric resistance (electric breakdown) and this new
type of electric conductivity also significantly affects the heat
transport through a NEM device. An intriguing question occurring
in this context is whether or not the similar shuttle instability
can be induced thermally at zero bias voltage applied to the
device. In other words --- is there a room for mechanically
induced thermal breakdown in NEM shuttle devices?

The electric force, which drives a charged movable quantum dot,
vanishes in the zero voltage limit, implying that the coupling
between mechanical and electronic degrees of freedom of the NEM
device disappears. No pumping energy can be extracted from an
electrically unbiased device. In what follows we will show that in
a magnetic shuttle \cite{kulinich1} the magnetic exchange force
can provide the necessary work to induce a mechanical instability.
Therefore a thermal breakdown in a magnetic unbiased shuttle
device can take place.

In a mechanically soft NEM shuttle device, where a quantum dot
(QD) is coupled by electron tunneling to source and drain
electrodes, an onset of a mechanical instability occurs when the
bias voltage exceeds a critical value known as the instability
threshold. When this happens a limit cycle of mechanical
oscillations is reached (self-oscillation regime) and a steady
state electrical current, which provides mechanical transportation
of charge, is established. This current, which typically exceeds
the tunnel current in the absence of a shuttle instability by
several orders of magnitude, is called an electrical shuttle
current (see Refs. [\onlinecite{kulinich1}] and
[\onlinecite{gorelik}]).

In an electric shuttle device both the energy source and the driving force
 are electrical in nature: the energy source is a bias voltage applied
between the leads and the driving force is the Coulomb force between
charges in the dot and the leads. In a magnetic shuttle device with
spin polarized electrons in the leads the energy source and the driving
force may have a different physical nature: the energy source could still be a bias
voltage while the driving force could be the magnetic (exchange) force between
electron spins in the dot and the magnetic leads. (For a nanometer size geometry
the exchange force may be as strong as the Coulomb force
\cite{mceuen}.) Therefore, it is in principle possible to
employ a thermal rather than an electrical energy source in a magnetic
shuttle device, i.e. to apply a temperature difference between
leads kept at the same chemical potential. In this paper we show
that such thermally induced magnetic shuttling is possible.

The action of the spin force is different from that of the Coulomb force. While the
Coulomb force tends to repel electrons transferred to the dot from
the lead they were injected from (Coulomb repulsion), the magnetic
force, caused by the spin of the injected electron, acts in the
opposite direction. \cite{footnote} Therefore the work done by the
magnetic force has the opposite sign compared to the work
performed by the Coulomb force.
Hence, the exchange force itself can not
pump energy into the
mechanical subsystem. However, if an external magnetic field $H$ perpendicular to the
magnetization in the leads is applied, this becomes possible. Such a field forces electrons in the dot to flip their
spins and the densities of spin-up and spin-down electrons in the
dot oscillate with a frequency determined by the magnetic
field. Therefore the direction of the
magnetic exchange force may change. As a result
it becomes possible to trigger a
shuttle instability in a magnetic device by applying an external magnetic field.

In what follows we will assume for simplicity that the magnetic leads
are fully spin-polarized and that their magnetizations are anti-parallel. Under this condition the electrical current is
blocked completely until ``spin flips" \cite{footnote2}
are induced by the external magnetic field (which is assumed to be
oriented perpendicularly to the magnetization of the leads). The
influence of a partial spin polarization on the shuttle instability was
considered in Refs. [\onlinecite{ilinskaya1}] and
[\onlinecite{ilinskaya2}].

The model we use to study a thermally induced magnetic shuttle is
sketched in Fig.~1. It is the standard shuttle device (see, e.g., Refs.
[\onlinecite{kulinich1}] and [\onlinecite{kulinich2}]), the only difference
being that a temperature drop $\delta T$ is applied to the leads instead
of an electrical bias voltage and that one is interested in the heat flow
$J_q$ in response to this temperature drop. The thermal resistance
$R_T$ can be defined in analogy with the electrical resistance as
$R_T=\delta T/J_q$. An exponential decrease in the thermal resistance
(thermal breakdown) is possible due to transduction of thermal
energy into the mechanical energy stored in the shuttle vibrations.

Below we will show that a mechanical shuttle instability
 occurs within
a finite interval of external magnetic fields strengths, $(H_{c1}<H<H_{c2})$.
The dependence of the upper, $H_{c2}$, and lower, $H_{c1}$, threshold
magnetic fields [which separate the shuttle and tunnel (outside
this interval) regimes of electron transport] on the large
temperature difference $\delta T$ close to the temperature $T$ of
the ``hot" lead is
the main result of our paper. The lower threshold field $H_{c1}$
is determined by the dissipation (friction) coefficient in the mechanical
subsystem. The friction coefficient
$\gamma_f=\gamma_0+\gamma_J(T)$ is the sum of a (phenomenological) friction coefficient $\gamma_0$
(see e.g. Ref.~\onlinecite{grabert})
and the intrinsic friction coefficient \cite{santandrea} $\gamma_J(T)$ induced
in our case by magnetic exchange forces and electron tunneling
between the dot and the leads at finite temperatures. We will call
this coefficient the {\it magnetic} friction coefficient. We will show
that the phenomenological friction coefficient can be neglected
for a mechanical system with a high quality factor. The main
contribution to magnetic friction is due to the hot electron
reservoir (we assume a large temperature difference). We show that
magnetic friction exists even in the absence of an external magnetic
field and that it is a non-monotonic function of temperature: it is
exponentially small at low temperatures, becomes
temperature-independent in the region $T\simeq\Gamma$ (where
$\Gamma$ is the characteristic energy of the dot-lead coupling) and
scales as $1/T$ in the high temperature limit. Since the pumping
of energy into the mechanical subsystem in our model is triggered by
the external magnetic field (the corresponding
 rate of increase of the oscillation amplitude in low magnetic fields being
proportional to $H^2$) a lower threshold magnetic field $H_{c1}$
with a nontrivial temperature dependence appears. As we have shown before
\cite{kulinich1} the mechanical
instability disappears in high magnetic fields. This is why an upper
threshold field $H_{c2}$ appears in the temperature driven shuttle as well.

Our calculations do not give any information about the low temperature
limit, ($\delta T\ll\Gamma$), since we use an
approximation where the thermal energy is large compared to the width of the energy levels on the dot
(sequential electron tunneling) and since in our model $\delta T=T$.
However, our previous results (see, e.g.,
[\onlinecite{fedorets}]) allows us to expect that the instability
occurs at temperature differences not smaller than a value of the
order of $\hbar\omega$. Therefore one may expect a non-monotonic dependence of the upper
threshold field on the temperature difference with a maximum at $\delta T\sim\Gamma$).

This paper is organized as follows. In Section II the model system
we use to discuss thermo-induced single-electron shuttling is introduced; an equation for
the reduced density operator of the QD and an equation of motion for the
classical coordinate of the dot are obtained. In Section III the domain where a magnetic shuttle
instability occurs in the adiabatic regime of electron transport is characterized.
In the concluding Section IV we highlight the main results obtained and discuss
possible applications of a temperature induced shuttle instability.

\section{Thermo-Induced Single-Electron Shuttle}

\begin{figure}
\centering
\includegraphics[width=0.85\columnwidth]{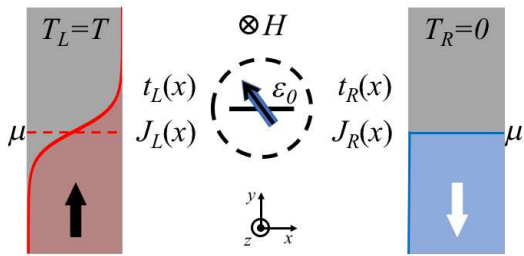}
\caption{Sketch of the nanomagnetic device studied: a movable
spin-degenerate single-level (with energy $\varepsilon_0$) quantum
dot is coupled by electron tunneling to two fully spin-polarized ferromagnetic
leads. The leads are kept at the same chemical potential $\mu$ but
at different temperatures $T_L=T$ and $T_R=0$. Here $t_L(x)$,
$t_R(x)$ and $J_L(x)$, $J_R(x)$ are dot position-dependent
tunneling amplitudes and exchange energies.  An external magnetic
field $H$ induces flips between the spin-up and spin-down states
on the dot.}\label{fig1}
\end{figure}

The system under consideration (see Fig.~\ref{fig1}) consists of a single-level quantum dot that is
 coupled by electron tunneling to two ferromagnetic
 electrodes (leads). The leads are fully spin-polarized with their magnetization pointing in opposite
directions. There is an external magnetic field $\overrightarrow
H$ in the gap between the source and drain leads, which is directed perpendicular to the magnetization in the leads.  We assume
that the leads are kept at equal chemical potentials
($\mu_L=\mu_R\equiv\mu$)
but at different temperatures $T_L\neq T_R$, so that a temperature
gradient $\delta T=T_L-T_R$ is applied to the system. The
proposed design of the electrodes (suitable for thermal transport
measurements as in Ref.~\onlinecite{dutta})  
allows one to maintain a temperature difference between the leads, while
keeping their chemical potentials equal. To simplify
calculations in what follows we will assume that $T_L=T$ and
$T_R=0$. It follows that in our system the temperature difference $\delta T=T$
and the mean temperature $T_m=T/2$ are not independent quantities.

The Hamiltonian of the system has three terms,
\begin{equation}\label{01}
{\cal\hat H}=\hat H_l+\hat H_d+\hat H_t.
\end{equation}
The Hamiltonian, $\hat H_l$, describes non-interacting electrons
in the electrodes,
\begin{equation}\label{02}
\hat H_l=\sum_{k,\kappa}\varepsilon_{k,\kappa}a_{k,\kappa}^\dag
a_{k,\kappa},
\end{equation}
where $a_{k,\kappa}^\dag (a_{k,\kappa})$ is the creation
(annihilation) operator of electron with momentum $k$ (energy
$\varepsilon_{k,\kappa}$) in the lead $\kappa=(L,R)$. The quantum
dot Hamiltonian reads ($\sigma=(\uparrow,\downarrow)=(+,-)$ is the
spin projection index),
\begin{eqnarray}\label{03}
&&\hat H_d=\sum_\sigma\varepsilon_\sigma c_\sigma^\dag c_\sigma
-\frac{g\mu_B H}{2}\left(c_\uparrow^\dag
c_\downarrow+c_\downarrow^\dag
c_\uparrow\right)+\nonumber\\&&+Uc_\uparrow^\dag c_\uparrow
c_\downarrow^\dag c_\downarrow+\hat H_v,\,\hat H_v=\frac{
p^2}{2m}+\frac{m\omega^2 x^2}{2},\label{003}
\end{eqnarray}
where $\varepsilon_\sigma=\varepsilon_0-(\sigma/2)J(x)$ is spin-
and position-dependent energy of quantum dot split levels
($\varepsilon_0$ is the level energy), $J(x)=J_L(x)-J_R(x)\simeq
J_0-\alpha x$ ($\alpha>0$ and we consider only small deviations,
$x$, of the dot center-of-mass coordinate from its equilibrium
position) is the coordinate-dependent exchange energy produced by the
ferromagnetic coupling between the dot and the leads, the operator
$c_\sigma^\dag (c_\sigma)$ creates (annihilates) an electron with
 spin projection $\sigma$ in the dot; $H$ is the external
magnetic field directed along the $z$-axis (see Fig.~1), $g$ is the
gyromagnetic ratio, $\mu_B$ is the Bohr magneton, $U$ is the
 Coulomb repulsion energy in the dot. Vibrations of the dot are
described by the harmonic-oscillator Hamiltonian $\hat H_v$
($m$ and $\omega$ are the mass and angular vibration frequency of
the dot). In what follows we will consider $x$ and $p$ as
classical time-dependent variables.

Tunneling of electrons between dot and leads is described
by the standard tunneling Hamiltonian
\begin{equation}\label{4}
\hat H_t=t_L(x)\sum_k c^\dag_\uparrow a_{k,L}+t_R(x)\sum_k
c^\dag_\downarrow a_{k,R}+\text{H.c.},
\end{equation}
where $t_\kappa(x)=t_\kappa \exp[\mp x/(4\lambda)]$ is the
tunneling amplitude, which has an exponential dependence on the dot
center-of-mass coordinate ($\lambda>0$ is the tunneling length, the signs
$``\mp"$ correspond to the left and right electrodes respectively).

The quantum description of the electron subsystem is based on the
assumption that the density matrix of the system can be factorized,
\begin{equation}\label{071}
\hat\rho(t)=\hat\rho_d\otimes\hat \rho_l,
\end{equation}
where $\hat\rho_l$ is the equilibrium density matrix (Gibbs distribution
function) of the leads. This assumption is always valid for
$T\gg\Gamma$ ($\Gamma$ is the tunnel coupling energy -- level width), when
sequential electron tunneling is the main process of electron
transport. In Eq.~(\ref{071}) $\hat\rho_d$ is the density matrix of the quantum dot
interacting with the magnetic leads.

In a general case one has to pay attention to the appearance of an implicit
time dependence of both the unperturbed Hamiltonian, $\hat
H_0=\hat H_l+\hat H_d$, and the tunneling Hamiltonian, $\hat H_t$,
due to the time dependence of the dot coordinate (and momentum),
$x(t),\,p(t)$. Therefore the derivation of the kinetic equations
in Ref.~\onlinecite{Fed1} requires some modifications.

The equation for the density operator (in units where $\hbar=1$),
\begin{equation}\label{051}
\frac{\partial \hat\rho(t)}{\partial t}+\imath\left[\hat H_0+\hat
H_t, \hat\rho(t)\right]=0,
\end{equation}
has the formal solution
\begin{equation}\label{06}
\hat\rho(t)=\hat\rho(t=-\infty)-\imath\int_{-\infty}^tdt'\hat
u(t,t')\left[\hat H_t(t'),\hat \rho(t')\right]\hat u^\dag (t,t'),
\end{equation}
where $\hat u(t,t')$ is the evolution operator of the unperturbed
Hamiltonian,
\begin{equation}\label{07}
\hat u(t,t')=e^{-\imath \hat H_l(t-t')}\hat u_d(t,t'),\,\hat
u_d(t,t)=1.
\end{equation}
In Eq.~(\ref{07}) $\hat u_d(t,t')$ is a dot evolution operator.
After substitution of Eqs.~(\ref{071}) and (\ref{06}) into Eq.~(\ref{051})
and tracing out the electronic degrees of freedom in the leads one
gets
\begin{widetext}
\begin{eqnarray}\label{08}
&&\frac{\partial \hat \rho_d(t)}{\partial t}+\imath\left[\hat
H_d,\hat \rho_d\right]=-\text{Tr}\int_{-\infty}^tdt' \left[\hat
H_t(t),e^{-\imath \hat H_l(t-t')} \hat u_d(t,t')\left[\hat
H_t(t'), \hat \rho(t')\right]e^{\imath \hat H_l(t-t')}\hat
u_d^\dag(t,t')\right].
\end{eqnarray}
\end{widetext}
The term on the r.h.s. of Eq.~(\ref{08}) has the sense of a collision
integral, $\hat I=\hat I_L+\hat I_R$, due to the interaction
between the dot and the leads. The kernel of this integral is
expressed through the function $K_\kappa(t,t')$  that can be
evaluated exactly in the wide-band approximation limit, when
one assumes that the density of states in the leads is energy
independent,
\begin{eqnarray}\label{09}
&&K_\kappa(t,t')=K_\kappa(t-t')\equiv K_\kappa(\tau)=\nonumber\\
&&=\sum_k e^{-\imath \varepsilon_{k,\kappa}\tau}
f(\varepsilon_{k,\kappa})=\frac{\imath\pi\nu_\kappa T_\kappa
e^{-\imath \mu \tau}}{\sinh\pi(T_\kappa \tau+\imath 0)}.
\end{eqnarray}
In Eq.~(\ref{09}) $\nu_\kappa=\text{const}, T_\kappa
(=\beta_\kappa^{-1})$ are the density of states and the
temperature (inverse temperature) in the lead $\kappa,\,
f(\varepsilon)$ is the Fermi-Dirac distribution function, $\mu$ is
the chemical potential. As stated above we restrict
ourselves to the case of zero temperature in the right lead,
$T_R=0$. Then using the well-known formula from the
theory of distribution functions,
\begin{equation}\label{0999}
\frac{e^{\imath\tau z}}{\tau-\imath 0}=\left\{
\begin{array}{l}
2\imath\pi\delta(\tau)\,, \,\hspace{0.5cm}z\to\infty,\\
0\,, \hspace{1.5cm}z\to -\infty, \\
\end{array}\right.\nonumber
\end{equation}
one readily gets the following expression for the collision integral $\hat I_R$
in the regime of non-resonant tunneling,
$(\varepsilon_0-\mu)/\Gamma_\kappa\gg 1$,
\begin{equation}\label{010}
\hat I_R=\Gamma_R(x)\left[c_\downarrow
\hat\rho_d(t)c^\dag_\downarrow-\frac{1}{2}\left[\hat\rho_d(t),
c^\dag_\downarrow c_\downarrow\right]_+\right],
\end{equation}
where $[\hat A,\hat B]_+=\hat A\hat B +\hat B\hat A$ is an
anticommutator and $\Gamma_\kappa(x)=2\pi\nu_\kappa t^2_\kappa(x)$
is the partial level width.

The reduced density operator $\hat\rho_d(t)$ acts in Fock space,
which in our case is the finite dimensional space of a single
electron level on the dot. Matrix elements of the density operator
are
\begin{eqnarray}\label{13}
&&\rho_0=\langle 0\vert\hat\rho_d\vert
0\rangle,\,\rho_\sigma=\langle
\sigma\vert\hat\rho_d\vert\sigma\rangle,\nonumber\\
&&\rho_{\sigma\sigma'}=\langle
\sigma\vert\hat\rho_d\vert\sigma'\rangle,\, \rho_2=\langle
2\vert\hat\rho_d\vert 2\rangle,
\end{eqnarray}
where $\vert \sigma\rangle=c_\sigma^\dag\vert 0\rangle, \vert
2\rangle=c_\uparrow^\dag c_\downarrow^\dag \vert 0\rangle$,
$\rho_\sigma\equiv\rho_{\sigma\sigma}$, $\sigma\neq\sigma'$.
In what follows we restrict ourselves to the
Coulomb blockade regime, $ U\gg T$. Under this condition the
doubly occupied state is forbidden, $\rho_2=0$.

In a classical description of the vibrational degrees of freedom,
Eq.~(\ref{003}), the Hamilton equations for the dot coordinate and
momentum take the form
\begin{eqnarray}\label{013}
&&\hspace{1cm}\frac{\partial x}{\partial t}=\text{Tr}\left\{\hat
\rho_d(t)\frac{\partial \hat H_d}{\partial
p}\right\}=\frac{p}{m},\\&&\hspace{-1cm}\frac{\partial p}{\partial
t}=-\text{Tr}\left\{\hat \rho_d(t) \frac{\partial \hat
H_d}{\partial x}\right\}=-m\omega^2x-\frac{\alpha}{2}
\left(\rho_\uparrow- \rho_\downarrow\right).
\end{eqnarray}
The oscillator coordinate $x(t)$ obeys the integro-differential
equation
\begin{equation}\label{19}
\frac{\partial^2x}{\partial t^2}+\omega^2
x=-\frac{\alpha}{2m}\left(\rho_\uparrow-\rho_\downarrow\right),
\end{equation}
where $\rho_{\uparrow,\downarrow}$ are functionals of coordinate,
$\rho_{\uparrow,\downarrow}=\rho_{\uparrow,\downarrow}\{x(t)\}$.

\section{Adiabatic regime of dot oscillations}

In the adiabatic limit $\omega\ll\Gamma_\kappa$ when evaluating
the collision integral $\hat I_L$ one can neglect the dependence
of coordinate on time. Then the evolution operator of the dot
takes the form,
\begin{equation}\label{190}
\hat u_d(t,t')=\exp\left[-\imath \hat H_d(t-t')\right].
\end{equation}
After straightforward calculations the collision integral $\hat
I_L$ in Eq.~(\ref{08}) can be represented in the form (we omit the
index "$d$" in $\hat \rho_d(t), \hat H_d$ and index "$L$" in
$\beta_L, T_L$):
\begin{widetext}
\begin{eqnarray}\label{11}
&&\hspace{5cm}\hat I_L=\frac{\Gamma_L(x)}{2}
\left[c_\uparrow^\dag\hat \rho(t) c_\uparrow+c_\uparrow
\hat\rho(t) c_\uparrow^\dag-\hat
\rho(t)\right]+\frac{\imath \Gamma_L(x)}{4}\times\nonumber\\
&&\hspace{-0.5cm}\times\left\{\int_{-\infty}^\infty d\tau
\frac{e^{\imath \beta \mu \tau}}{\sinh\pi\tau}c_\uparrow
e^{-\imath\beta\hat
H\tau}\left[\rho(t-\beta\tau),c_\uparrow^\dag\right]_+
e^{\imath\beta\hat H\tau} +\int_{-\infty}^\infty d\tau
\frac{e^{-\imath \beta \mu \tau}}{\sinh\pi\tau}c_\uparrow^\dag
e^{-\imath\beta\hat H\tau}\left[\rho(t-\beta\tau),
c_\uparrow\right]_+ e^{\imath\beta\hat H\tau}-\text{H.c.}\right\}
\end{eqnarray}
\end{widetext}
(here $\tau$ is the dimensionless integration variable).
In Eq.~(\ref{11}) the singular integrals are understood in the
sense of the principal value. In the limit of high temperatures,
$\Gamma_L \ll T$, one can neglect the retardation effects and
replace $\rho(t-\beta\tau)\rightarrow \rho(t)$ in Eq.~(\ref{11}).

From Eqs.~(\ref{08}), (\ref{010}), (\ref{11}) one gets the
following system of equations for the matrix elements of the
density operator (note, that  the Hamiltonian $\hat H_d$ is not
diagonal in $\sigma$-representation, but it can be easily
diagonalized by unitary transformation):
\begin{eqnarray}
&&\frac{\partial\rho_0}{\partial
t}=\Gamma_L(x)\left(1-f_+\right)\rho_\uparrow
-\Gamma_L(x)f_+\rho_0+
\Gamma_R(x)\rho_\downarrow-\nonumber\\
&&-\Upsilon_1(x) \left(\rho_0+\rho_\uparrow\right)
-\Upsilon_2(x)\left(
\rho_{\uparrow\downarrow}+
\rho_{\uparrow\downarrow}^\ast\right),\label{14}\\
&&\frac{\partial\rho_\uparrow}{\partial t}
=-\Gamma_L(x)\left(1-f_+\right) \rho_\uparrow-\imath \Omega_H
\left(\rho_{\uparrow\downarrow}
-\rho_{\uparrow\downarrow}^*\right)+\\
&& +\Gamma_L(x)f_+\rho_0+\Upsilon_1(x)
\left(\rho_0+\rho_\uparrow\right)+\Upsilon_2(x)\left(
\rho_{\uparrow\downarrow}+\rho_{\uparrow\downarrow}^\ast
\right),\nonumber\\
&&\frac{\partial \rho_\downarrow}{\partial
t}=-\Gamma_R(x)\rho_\downarrow+ \imath \Omega_H
\left(\rho_{\uparrow\downarrow}-\rho_{\uparrow\downarrow}^*\right),\\
&&\frac{\partial \rho_{\uparrow\downarrow}}{\partial t}=\imath
J(x)\rho_{\uparrow\downarrow}-\imath \Omega_H\left(\rho_\uparrow-
\rho_\downarrow\right)-\frac{\Gamma_L(x)}{2}\left(1-f_+\right)
\rho_{\uparrow\downarrow}\nonumber\\
&& -\frac{\Gamma_R(x)}{2} \rho_{\uparrow\downarrow}-
\frac{\Upsilon_1(x)}{2}\rho_{\uparrow\downarrow}
+\Upsilon_2(x)\left(\rho_0+\rho_\uparrow\right),\label{15}
\end{eqnarray}
where $\Omega_H=g\mu_B H/2$ and

\begin{eqnarray}\label{18}
&&\Upsilon_1(x)=f_-\frac{J(x)\Gamma_L(x)}
{\sqrt{J^2(x)+4\Omega_H^2}},\\&&\Upsilon_2(x)=f_-
\frac{\Omega_H\Gamma_L(x)}{\sqrt{J^2(x)+4\Omega_H^2}},\\
&&f_\pm=\frac{f(E_-)\pm f(E_+)}{2},\\
&& E_\pm=\varepsilon_0\pm\frac{\sqrt{J^2(x)+4\Omega_H^2}}{2}.
\end{eqnarray}

To simplify the problem we consider the symmetric quantum dot,
$J_0=0,\, \Gamma_L(0)=\Gamma_R(0)=\Gamma$. We are interested in
the conditions when the stationary position of the dot ($x=0$) is
not stable. In this case it is sufficient to consider small
deviations $x/\lambda\ll 1$ and to linearize the coordinate
dependence of $\Gamma_\kappa(x)\simeq\Gamma\left(1\mp x/2\lambda
\right)$.

At first we solve the problem in the high temperature limit,
$\beta\rightarrow 0\, (f_-=0, f_+=1/2)$. It is convenient to
rewrite the system, Eqs.~(\ref{14}) -- (\ref{15}), in new
variables,
\begin{equation}\label{20}
R_{1,2}=\rho_\uparrow\pm\rho_\downarrow,\,
R_3=-\imath\left(\rho_{\uparrow\downarrow}-\rho_{\uparrow
\downarrow}^*\right),\, R_4=\rho_{\uparrow\downarrow}+
\rho_{\uparrow\downarrow}^*.
\end{equation}
In what follows we will assume that the dimensionless parameter
$\tilde \alpha= \alpha/(m\lambda \omega^2)$ is small, $\tilde
\alpha\ll 1$. Since we study small vibrations of the dot, one can
solve the system, Eqs.~(\ref{14}) -- (\ref{15}) by perturbations,
$\vert R\rangle= \vert R^{(0)}\rangle+\vert R^{(1)}\rangle+...$,
where $\vert R\rangle=\left(R_1, R_2, R_3\right)^T$ (note, that
the equation for $R_4$ is decoupled from the other equations and
it is not relevant). In zero order of perturbation theory one gets
\begin{eqnarray}\label{21}
&&\langle R^{(0)}\vert =
\frac{1}{2\Delta}\left(\frac{3\Gamma^2}{4}+4\Omega_H^2,
\frac{3\Gamma^2}{4}, -2\Gamma \Omega_H\right),\nonumber\\
&&\hspace{2cm}\Delta=\frac{3\Gamma^2}{4}+5\Omega_H^2.
\end{eqnarray}
In the first order of perturbation theory the equation for $\vert
R^{(1)}\rangle$ takes the form
\begin{equation}\label{22}
\frac{\partial \vert R^{(1)}\rangle}{\partial t}=\hat A\vert
R^{(1)}\rangle+\frac{\Gamma }{2\lambda}x(t)\vert g\rangle,
\end{equation}
where
\begin{equation}\label{23}
\hat A=-\frac{\Gamma}{4}\left(\begin{array}{ccc}
  5 & -1 & 0 \\
  1 & 3 & -8\Omega_H/\Gamma \\
  0 & 8\Omega_H/\Gamma & 3 \\
\end{array}\right),\,\vert g\rangle=\frac{\Omega_H}{4\Delta}\left(\begin{array}{c}
  -8\Omega_H \\
  0 \\
  \Gamma\\
\end{array}\right).
\end{equation}
Substituting the solution of Eq.~(\ref{22}) into the r.h.s. of
Eq.~(\ref{19}) we derive the desired equation for single electron
shuttle coordinate
\begin{equation}\label{24}
\frac{\partial^2x}{\partial t^2}+\omega^2 x=-\frac{\alpha
\Gamma}{4\lambda m}\int_0^\infty d\tau \langle e_0\vert e^{\hat A
\tau} \vert g\rangle x(t-\tau),
\end{equation}
where $\langle e_0 \vert=(0, 1, 0)$.

In the adiabatic limit $\omega\ll \Gamma$ one can expand
$x(t-\tau)\simeq x(t)-\tau \dot{x}(t)$. We see that the
electro-mechanical coupling results in (small) additive
renormalization of vibrational frequency $\omega$ and the
appearance of damping (or pumping) term $\gamma\dot{x}$ in the
mechanical equation, where the coefficient $\gamma(\Gamma,\Omega_H)$
reads
\begin{eqnarray}\label{25}
&&\gamma(\Gamma,\Omega_H)=-\frac{\alpha \Gamma}{4\lambda m}\int_0^\infty
d\tau \tau \langle e_0\vert e^{\hat A \tau} \vert
g\rangle=\nonumber\\&&\hspace{1.3cm}=\frac{\hbar\alpha \Gamma
\Omega_H^2}{4\lambda m \Delta^3}\left(\Omega_H^2-\frac{7\Gamma^2}{4}\right)
\end{eqnarray}
(we restored the dimension in the last formula). It is easy to
find from Eq.~(\ref{25}) that in weak magnetic fields,
\begin{equation}\label{26}
\left\vert\frac{ g\mu_B H}{2}\right\vert<\frac{g\mu_B
H_{c2}}{2}=\frac{\sqrt{7}}{2}\Gamma,
\end{equation}
the shuttle instability occurs. Note, that the increment
$r(\Gamma,\Omega_H)=-\gamma(\Gamma,\Omega_H)/2$ of the exponential
growth of shuttle oscillations amplitude in the limit $\omega\ll\Gamma$ does not depend on
the dot frequency $\omega$.

For finite temperatures the calculations are similar to the
previous ones but they are more lengthy. For simplicity we
restrict ourselves to the case of relatively large magnetic
fields, $\vert \Omega_H\vert\gg\alpha \lambda$. Under this condition for
the damping (pumping) coefficient one gets the expression
\begin{equation}\label{27}
\gamma_T(\Gamma,\Omega_H)=-\frac{\alpha\Gamma}{8\lambda m}\int_0^\infty
d\tau \tau \langle e_1\vert e^{\hat A_T \tau} \vert g_1\rangle,
\end{equation}
where $\langle e_1\vert =(0, 1, 0, 0)$ and
\begin{equation}\label{28}
\hat A_T=-\frac{\Gamma}{2}\left(\begin{array}{cccc}
  2+f_+ & f_+ & 0 & -f_- \\
  -f_+ & 2-f_+ & 2\Omega_H/\Gamma & f_- \\
  0 & -2\Omega_H/\Gamma & 2-f_+ & 0 \\
  f_- & -f_- & 0 & 2-f_- \\
\end{array}\right),
\end{equation}
\begin{eqnarray}\label{g1}
&&\vert g_1\rangle=\frac{4}{\Delta_T}\left(\begin{array}{cccc}
-4\Omega_H^2\left[f_+(2-f_+)+f_-^2\right]\\0\\
\Gamma \Omega_H \left[f_+(2-f_+)+f_-^2\right]f_+\\
\left[\Gamma^2(2-f_+)+8\Omega_H^2\right]f_-\end{array}\right),\\
&&\Delta_T=\Gamma^2(2-f_+)^2+4\Omega_H^2(4-f_+^2+f_-^2).
\label{DeltaT}
\end{eqnarray}
The shuttle instability condition is given by the inequality
\begin{equation}\label{inequality}
C_1\left(\Omega_H/\Gamma\right)^4- C_2\left(\Omega_H/\Gamma\right)^2+C_3<0,
\end{equation}
where
\begin{eqnarray}\label{coeff}
&&C_1=2f_+^3(2-f_+)^3-f_+f_-^2(2-f_+)^2(4-5f_+)-\nonumber\\
&&-4f_-^4(2-f_+)(1-f_+)-f_-^4(4-f_-^2),\\
&&C_2=\frac{2-f_+}{2}\left[f_+^2(2-f_+)^2(4-f_+)-\nonumber\right.\\
&&\left.-f_-^2(2-f_+)(4-f_+)(1-2f_+)+f_-^4(5-f_+)\right],\\
&&C_3=\frac{(2-f_+)^3 (4-f_+)f_-^2}{16}.\label{251}
\end{eqnarray}
As a consequence, the shuttle instability region is defined by the
(transcendental) relation
\begin{equation}\label{252}
\Omega_{Hc1}^2<\Omega_H^2<\Omega_{Hc2}^2,
\end{equation}
where
\begin{equation}\label{253}
\Omega_{Hc1(c2)}^2=\Gamma^2\frac{C_2\mp\sqrt{C_2^2-4C_1C_3}}{2C_1}.
\end{equation}

The lower critical magnetic field $\Omega_{Hc1}$ lies outside the
range of applicability of our calculations. (We neglected the
amplitude of shuttle oscillations compared to $\Omega_H/\alpha$.)
Physically the existence of the lower critical magnetic field can
be easily explained. Even in the absence of an external magnetic
field (and in the absence of phenomenological friction) at finite
temperature there is dissipation in the mechanical subsystem
induced by magnetic forces and back-tunneling of electrons to the
hot lead. The corresponding friction coefficient $\gamma_J(T)$ (in
what follows we will call it magnetic friction) can be estimated
from simple physical considerations. Magnetic friction appears due
to a finite work performed by magnetic driving force along the
closed trajectory of oscillating quantum dot and therefore it is
proportional to the coordinate derivative of Fermi distribution
function $f[\varepsilon_{\uparrow}(x)-\mu]$. Since magnetic force
is nonzero only when the electron level is occupied, magnetic
friction depends on the dot-lead coupling energy $\Gamma$. By
taking into account retardation effects \cite{footnote3} this
contribution to magnetic friction is represented by a factor
$\Gamma/(\Gamma^2+(\hbar\omega)^2)$. As the result friction
coefficient takes the form
\begin{equation}\label{rJ}
r_J(T)\sim
-\frac{\hbar\alpha^2}{m}\frac{\Gamma}{\Gamma^2+(\hbar\omega)^2}\frac{1}{T}
\cosh^{-2}\left(\frac{\delta\varepsilon}{2T}\right)\,,
\end{equation}
where $\delta\varepsilon=\varepsilon_0-\mu$. Note that friction
coefficient is defined as $\gamma_J(T)=-2r_J(T)$. The calculation
of the decrement of shuttle vibrations in the absence of external
magnetic field ($\Omega_H=0$) by using Eqs.~(15), (18--25) leads
to Eq.~(\ref{rJ}) with the numerical prefactor 1/32. We see that
in high-T limit $T\gg\delta\varepsilon$ magnetic friction is
decreased with the growth of temperature. At temperatures
$\Gamma\ll T\ll\delta\varepsilon$ dissipation is exponentially
small, $r_J\propto\exp(-\delta\varepsilon/T)$. Our calculations
are not valid at temperatures $T\ll\Gamma$ where resonant electron
tunneling takes place. However it is evident from physical
considerations that dissipation vanishes when $T\rightarrow 0$.
Anomalous temperature behavior of $r_J(T)$ is a specific feature
of magnetic dissipation which takes maximum value at
$T\sim\Gamma\sim\delta\varepsilon$ and it vanishes in the limits
of both small and high temperatures.

The shuttle instability appears when the increment of exponential
growth of dot oscillations amplitude exceeds the decrement
$\gamma_J(T)/2$. For small magnetic fields, $\Omega_H\to 0$, the
increment reads
\begin{equation}\label{increment}
r(\Gamma,\Omega_H\rightarrow 0)\simeq\frac{14\hbar\alpha}{3m\lambda\Gamma^3}\Omega_H^2\,.
\end{equation}
Therefore, by comparing Eqs. (\ref{rJ}) and (\ref{increment}) we
can estimate the lower critical magnetic field in the high-$T$ limit as
\begin{equation}\label{lower}
\Omega_{Hc1}\simeq 0.1\sqrt{\frac{\alpha\lambda}{T}}\Gamma\,.
\end{equation}
The phenomenological friction coefficient $\gamma_0=\omega/Q$ can
be neglected in comparison with the optimal intrinsic friction
coefficient $\gamma_J(T)$ if the quality factor $Q$ of the
mechanical subsystem is sufficiently large. We estimate the
minimal quality factor required to be $Q_{min}\sim 10^2 \div 10^3$
for $\Gamma\sim\hbar\omega\sim 1\, \text{meV}$, $J_L(0)\sim
J_R(0)\sim 10\, \text{meV}$, values taken from experimental work
[\onlinecite{mceuen}, \onlinecite{mceuen2000}] on $C_{60}$-based
molecular transistors.

\begin{figure}
\centerline{\includegraphics[width=0.75\columnwidth]{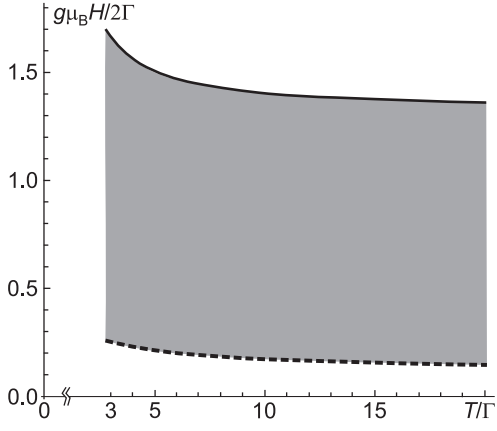}}
\caption{The lower ($H_{c1}$, dashed curve) and upper ($H_{c2}$,
full curve) threshold magnetic fields plotted as functions of
normalized temperature, $T/\Gamma$, for
$\delta\varepsilon/\Gamma=2$ in the adiabatic regime,
$\omega\ll\Gamma/\hbar$. These fields define the border between
the shuttle regime ($H_{c1}<H<H_{c2}$, shaded region) and the
vibronic regime (outside this interval). For $T\to\infty$ the
lower threshold field $H_{c1}$ saturates to a constant value
determined by the phenomenological friction coefficient
$\gamma_0$.}\label{fig2}
\end{figure}

The shuttle instability domain (shaded region in Fig.~\ref{fig2})
is plotted in $T/\Gamma, g\mu_B H/2\Gamma$ parameter space for
$\delta\varepsilon/\Gamma=2$. The shuttle domain is shown only for
$T\geq\Gamma$ because we solved the problem in the perturbation
theory in small parameter $\beta\Gamma\ll 1$. Although our
calculations are not valid at low temperatures, at $T\rightarrow
0$ the increment $r(\Gamma,\Omega_H)$ behaves as
$r\sim\exp\left[-2\beta\left(
\delta\varepsilon-\Omega_H\right)\right]$. Exponential smallness
of $r_T(\Omega_H)$ for $T\rightarrow 0$ is physically reasonable
result.  In the high temperature limit, $T\to\infty$, we return to
the result of Eq.~(\ref{26}) for $H_{c2}$. Leaving the next term in expansion in small parameter
$\beta\delta\varepsilon$, we obtain such an asymptotic
behaviour of the critical magnetic field at large temperatures,
\begin{equation}\label{270}
\frac{g\mu_B H_{c2}}{2}\simeq\frac{\sqrt{7}}{2}\left[1+
\frac{2}{7}\frac{\delta\varepsilon}{T}\right]\Gamma.
\end{equation}

In adiabatic limit ($\omega\ll\Gamma$) we used the evolution
operator of the dot, $\hat u_d(t,t')$, in the form of
Eq.~(\ref{190}). However the criterion of validity of the
expression for the evolution operator in this form for magnetic
shuttle is not equivalent to condition $\omega\ll\Gamma$. In fact,
the analysis shows that the criterion of the validity of
Eq.~(\ref{190}) is ($ \Omega_H,\Gamma \neq 0$)
\begin{equation}\label{30}
\frac{\omega}{\Gamma}\frac{\alpha\lambda
\Omega_H}{\Omega_H^2+(\alpha\lambda)^2}\ll 1.
\end{equation}
Therefore, in the limit $\alpha\lambda/\Omega_H\ll 1$ the ratio
$\omega/\Gamma$ can take large values ($\omega/\Gamma\geq 1$)
without violation of adiabaticity of mechanical
motion.

Besides in our consideration we assume that the parameter $\tilde
\alpha=\alpha/(m\lambda\omega^2)$ is small, $\tilde \alpha\ll 1$.
When both inequalities are taken into account one gets upper bound for frequencies
\begin{equation}\label{upb}
 \omega\ll\omega_m =\left[\frac{\Omega_H\Gamma}{ m
\lambda^2}\right]^{1/3}
\end{equation}
when the evolution operator can be considered in the form corresponding
to adiabatic motion.

When the conditions of Eqs.~(\ref{30}), (\ref{upb}) are fulfilled
one can use the system of kinetic equations, Eqs.~(\ref{14}) --
(\ref{15}), and to analyze the behavior of the system at high
frequencies similar to the previous calculations. As a result the
shuttle instability at frequencies higher than $\Gamma$ is defined
by the inequality
\begin{equation}\label{35}
\vert \Omega_H\vert/\omega <C(\Omega_H,T),
\end{equation}
where
\begin{eqnarray}\label{36}
&&C(\Omega_H,T)=\nonumber\\
&&\hspace{-0.5cm}\sqrt{\frac{(2-f_+)(4-f_+)(2f_+^2-f_-^2)+f_-^2(2f_+-f_-^2)}
{4(f_+^2-f_-^2)[2f_+(2-f_+)+f_-^2]}}.
\end{eqnarray}
(In formulas (\ref{35}) -- (\ref{36}) we assumed
$\omega\gg\Gamma$.)

The shuttle instability domain plotted in $T/\hbar\omega, g\mu_B
H/2\hbar\omega$ parameter space has the same form as the shuttle
instability domain at small frequencies plotted in Fig.~2.

We would like to note here another interesting fact. In the limit
$T\rightarrow\infty$ the problem under consideration can be solved
exactly for arbitrary relationship between the model parameters
$\Omega_H,\Gamma, \omega$. Physically the considered infinite temperature
limit is realized for temperatures $T\gg\text{max}(\hbar\omega,
\Gamma)$. In this limit the kernel of collision integral in
Eq.~(\ref{08}) can be replaced by $\delta$-function and the
integro-differential equation for density operator becomes local
in time. Indeed at $T\rightarrow\infty$
\begin{equation}\label{29}
\lim\limits_{T\to\infty}\frac{T}{\sinh\pi\left(T\tau+\imath0\right)}=-
\imath\delta(\tau)
\end{equation}
and the function $K_L(t,t')$ that defines the kernel of collision
integral, Eq.~(\ref{08}), is reduced to $K_L(t,t')=\pi\nu_L
\delta(t-t')$. As a consequence the evolution operator of the dot
is trivial (unit operator) and the system of kinetic equations for
the components of the density operator has a Markovian form. It is
obvious that in the limit
$\Gamma<\omega\ll\omega_m$
 this system coincides with Eqs.
(\ref{14}) -- (\ref{15}) for adiabatic case in the limit
$T\rightarrow \infty$. Therefore the dot dynamics is described by
Eq.~(\ref{24}) and the criterion of shuttle instability (the range
of magnetic field) for high frequencies is
\begin{equation}\label{301}
\left\vert\frac{ g\mu_B H}{2}\right\vert<\frac{g\mu_B
H_{c2}}{2}=\frac{\sqrt 7}{2}\hbar\omega.
\end{equation}
This result is in agreement with Eqs.~(\ref{35}), (\ref{36}).

The increment $r(\Gamma,\Omega_H,\omega)$ of the exponential growth of shuttle center of
mass coordinate in the limit of high temperatures (note that in our model
$\delta T=T$) takes the form (we restore the dimensions)
\begin{equation}\label{311}
r(\Gamma,\Omega_H,\omega)=\frac{\tilde \alpha\Gamma^3\Omega_H^2 }{8\hbar\Delta
D}\left[\frac{7}{4}\left(\hbar\omega\right)^2-\Omega_H^2\right],
\end{equation}
where $\Delta$ is defined by Eq.~(\ref{21}) and
\begin{equation}\label{32}
D=\left[\left(\hbar\omega\right)^2-4\Omega_H^2\right]^2
+\left(\frac{\Gamma}{\hbar\omega}\right)^2\left[\frac{11}{4}
\left(\hbar\omega\right)^2-5\Omega_H^2\right]^2.
\end{equation}
The maximal value of the increment is reached at $\Omega_H=\hbar\omega/2$
when $r_{max}(\Gamma,\Omega_H=\hbar\omega/2,\omega)=\tilde\alpha
\Gamma/60\hbar$.

\section{Conclusions}

We have shown that in a magnetic shuttle structure
\cite{kulinich1} a temperature gradient between the leads can
trigger a shuttle instability, which leads to an exponential
growth of the amplitude of shuttle oscillations, even in the
absence of a voltage bias. This leads to a ``mechanically
supported thermal breakdown" in the form of an exponential growth
of the heat current (as well as of the electrical current) through
the device. In our model \cite{gorelik} of fully (and oppositely)
spin polarized electrons in the leads a spin blockade prevents a
current to flow in the absence of an external magnetic field.
Lifting the spin blockade by applying such a field results in a
shuttle instability if the field strength exceeds a certain
threshold value, $H_{c1}$, determined by the amount of dissipation
in the mechanical subsystem. When the leads are kept at finite
temperatures, there is an intrinsic dissipation mechanism
\cite{santandrea} (``magnetic friction"; independent of magnetic
field for low field strengths) caused by the magnetic force and
the exchange of electrons between the quantum dot and the leads.
In addition there is phenomenological friction, which can be
neglected if the quality factor of the mechanical subsystem is
large enough. The amount of magnetic friction in our model is
determined by the temperature of the ``hot" lead, the level energy
$\delta\varepsilon=\varepsilon_0-\mu$, the dot-lead coupling
energy $\Gamma$, and the dot vibration frequency. In the general
case of an asymmetric junction, $\Gamma_L\neq\Gamma_R$, and
nonzero temperatures in both leads, $T_L\neq T_R\neq 0$, the
magnetic friction is the sum of contributions produced by each
lead. We predict that a specific feature of the magnetic friction
is its anomalous temperature dependence. It vanishes in the limits
of small and high temperatures and attains a maximum value at
temperatures $T\sim\Gamma\sim\delta\varepsilon$. No shuttle
instability occurs in such high magnetic fields, $H>H_{c2}$,  that
the spin-flip time exceeds the characteristic time scale
determined by the maximum of the mechanical $(\sim\omega^{-1})$ or
electronic $(\sim\hbar/\Gamma)$ time scales. For sequential
electron tunneling  $H_{c2}$ saturates at $T\gg\Gamma$ and
slightly increases with the decrease of temperature (see Fig.~2).

It is useful to qualitatively discuss the dependence of
$H_{c1}$ on the temperature difference between the
two heat reservoirs when they are held at almost equal temperatures, $\delta T\ll T$. In
this case the rate of increase, $r$, of the amplitude
of the shuttle oscillations after an instability has occurred has a linear dependence on $\delta T$ (for
an electric shuttle the dependence of the corresponding rate on
bias voltage $V$ and model parameters, $r\propto V\Gamma$, was
calculated in Ref.~\onlinecite{Fed}). If $T_L\simeq T_R\gg\Gamma$
our approach (using the density operator method) is valid and
from physical considerations one can deduce that the rate of energy pumping
is proportional to $H_{c1}^2\delta T$. The temperature dependence of the
magnetic friction is still determined by an equation
similar to Eq.~(\ref{rJ}) and therefore the friction coefficient $\gamma_J(T)\propto 1/T$,
where $T$ is the average temperature. We see that now
$H_{c1}\propto 1/\sqrt{T\delta T}$ and that it is much larger than the
corresponding field calculated for $\delta T\simeq T$. Therefore, it may be an unrealistic proposition to use
such high values of the static external magnetic
field in experiments.


The exponential increase of the amplitude of the center-of-mass oscillations of the dot saturates when the energy pumped into the
dot vibrations equals the energy dissipated by the magnetic
friction. From a general point of view our device works as a
spintronic quantum heat engine \cite{anders}. A spin-polarized
(spin-``up") electron tunnels from the hot lead to the vibrating
quantum dot and for a certain time it is localized in the dot. In
the absence of an external magnetic field the only further dynamics
of the spin-up electron allows it to tunnel back to the ``source" electrode.
In this case the work done by the exchange forces results in the dot motion being damped.

An external magnetic field induces coherent electron spin dynamics
in the dot (spin-up/down oscillations) and therefore
a new channel of electron tunneling (from the dot to the ``drain" lead) is
opened. This
process results in positive work being done by
the exchange forces, which amounts to pumping energy from the hot
lead to the mechanical motion of the quantum dot (the device becomes a spintronic
single-electron heat engine). Note that the transformation of heat
into mechanical energy in our device is carried out by strongly
nonequilibrium and nonlinear processes. The shuttle instability is an
intrinsically threshold phenomenon (there is a minimum temperature
difference $\delta T\sim\hbar\omega$ for which a mechanical instability
can occur). Therefore it can not be described by
thermoelectric coefficients obtained using linear response theory.

We speculate that the predicted phenomenon of a mechanically induced thermal breakdown could
find useful applications in spintronic devices, when it is essential to avoid high temperature gradients on a chip.

\section*{Acknowledgments}

The authors thank L.Y.~Gorelik for useful discussions. O.A.I.,
S.I.K., and I.V.K. acknowledge financial support from NAS Ukraine
(Grant \textnumero  4.17-H and Scientific Programme
1.4.10.26/$\Phi$-26-3). This publication is partly (I.V.K.) based
on the research provided by the grant support of the State Fund
for Fundamental Research (Ukraine, project \textnumero
$\Phi$76/33683). This work was supported by the Institute for
Basic Science in Korea (IBS-R024-D1). R.I.S. and M.J. acknowledge
support  from the Swedish Research Council (VR). I.V.K. and R.I.S.
acknowledge hospitality of PCS IBS, Daejeon (Korea).


\begin{thebibliography}{99}
\bibitem{footnote}
Nonzero magnetization of a lead
implies spin-split electron energy levels, with majority-spin electrons
occupying the lower split level.
The transfer of such an electron from the lead to the dot diminishes the magnitude of the energy
split and therefore increases the
total energy of the electron. Hence, the magnetic force on the dot, which
acts to lower the energy of the system, will be
oriented towards the magnetic lead that the electron tunneled from. This
interaction is ferromagnetic irrespective of the nature of the
exchange interaction (ferromagnetic or antiferromagnetic) in the
bulk leads.
\bibitem{footnote2}  Here the term ``spin flips" does not
refer to any spin relaxation mechanism but to spin flips caused by
the coherent spin dynamics (spin-up/down oscillations) in a
magnetic field. The rate of these spin flips is much higher than
those related to spin relaxation processes.
\bibitem{footnote3}
Note that in our
approximation the equation for the dot density matrix is local;
retardation effects are related to the nonlocality of the
effective force in the equation of motion for the oscillator
coordinate [see Eq.~(\ref{24})].
\end{thebibliography}
\end{document}